\documentclass[aps,prl,groupedaddress,superscriptaddress,twocolumn,10pt,longbibliography]{revtex4-2}
\pdfoutput=1 
\usepackage[utf8]{inputenc}
\usepackage{hyperref}
\usepackage{graphicx}
\usepackage{orcidlink}
\usepackage{booktabs}
\usepackage{tabularx}

\usepackage{siunitx} 
\newcommand{\omatt}[1]{}

\usepackage{amssymb}
\usepackage{physics}
\usepackage{mathtools}
\usepackage{lipsum}

\DeclareSIUnit\bohr{\text{\ensuremath{a_\textup{0}}}}

\begin{document}

\title{Microscopic Rydberg electron orbit manipulation with optical tweezers}

\author{Homar Rivera-Rodr\'iguez \orcidlink{0009-0007-6454-7063}}
\affiliation{Max-Planck Institut für Physik komplexer Systeme, Nöthnitzer Stra\ss e 38, 01187 Dresden, Germany}

\author{Matthew T. Eiles\orcidlink{0000-0002-0569-7551}}
\affiliation{Max-Planck Institut für Physik komplexer Systeme, Nöthnitzer Stra\ss e 38, 01187 Dresden, Germany}
\affiliation{Department of Physics and Astronomy, Purdue University, West Lafayette, IN 47907, USA}

\author{Tilman Pfau\orcidlink{0000-0003-3272-3468}}
\affiliation{5. Physikalisches Institut and Center for Integrated Quantum Science and Technology, Universität Stuttgart, Pfaffenwaldring 57, 70569 Stuttgart, Germany}

\author{Florian Meinert\orcidlink{0000-0002-9106-3001}}
\affiliation{5. Physikalisches Institut and Center for Integrated Quantum Science and Technology, Universität Stuttgart, Pfaffenwaldring 57, 70569 Stuttgart, Germany}

\date{\today}

\begin{abstract}
Laser cooling and trapping of atomic matter waves in optical potentials has enabled
rapid progress in quantum science, particularly when combined with Rydberg excitation of the
atoms to induce long-range interactions. Here, we propose the local manipulation and spatio-temporal sculpting
of the electronic matter wave of a Rydberg atom by a laser field focused so that its beam width
is smaller than the Rydberg electron orbit. 
We compute the electronic eigenstates in the presence of a sharply focused Gaussian laser beam, and find strong Rydberg state mixing leading to large kilo-Debye dipole moments.
These can be modulated with high bandwidth controlled by the local tweezer intensity. 
Oscillations in the position-dependent level shifts, analogous to the potential wells allowing ultralong-range Rydberg molecules to form, provide opportunities for eccentric radial trapping of the Rydberg electron via ponderomotive forces acting on sub-orbital length scales.
\end{abstract}

\maketitle

Electron orbitals are central to the modern understanding of how matter is composed, from the electronic shell structure within an atom to the hybridized orbitals forming chemical bonds in molecules and the delocalized orbitals determining conduction in the solid state  \cite{Schroedinger1926, Heitler1927, Bloch1929}. Their local control and manipulation in molecules or materials is a challenging task, typically requiring atomic-scale resolution instruments such as tunneling or atomic force microscopes \cite{Giessibl2003,Wang2024}. However, electronic orbitals need not remain microscopic: in Rydberg atoms, the typical size of the electronic wave function can extend up to a few microns \cite{Gallagher1994,Balewski2013,Kleinbach2018,Dunning2009}. This opens an entirely different approach for the local manipulation of electron orbitals via optical microscopy techniques.
In particular, single-atom control enabled by tightly focused optical tweezers appears especially well-suited for this purpose, as tweezer beams can be focused, positioned, and even dynamically rearranged at sub-micron scale \cite{Browaeys2020,Kaufman2021,Duta2000,Younge2010,Barredo2020,Holzl2024}.

In this Letter, we propose and theoretically investigate the local manipulation of and spatio-temporal control over electronic orbitals via an optical tweezer piercing a giant Rydberg atom. 
We classify the field-controlled electronic eigenstates into two varieties: one composed of low-angular momentum states which possess sizeable dipole moments stemming from the laser perturbation, and the second containing highly dipolar orbitals marked by strong electron localization in the degenerate hydrogenic Rydberg states. The critical control parameter $\eta$ is the ratio of the laser waist to the Rydberg orbit, with small $\eta$ producing oscillatory level shifts and asymmetric electron orbitals reminiscent of ultralong-range Rydberg molecules, except that here the tweezer beam plays the role of a localized perturbation rather than a ground-state atom \cite{Greene2000,Shaffer2018,Eiles2019,Dunning2024}.
However, optical tweezer interactions provide a much higher degree of tunability for electronic orbital control due to the flexibility of the deployed light fields, which also enables fast modulation in time. We argue that the adiabatic eigenstates and their large dipole moments can be driven at MHz-scale bandwidth to realize a locally controlled atomic-scale Hertzian dipole. While the eigenstates are readily accessible via spectroscopy, the driven dipoles could be sensed by nearby Rydberg atoms serving as resonant receivers \cite{Adams2020}. Moreover, although the tweezer repels the Rydberg electron, we find that there exist deep local minima in the adiabatic potentials which enable trapping of the Rydberg atom through sub-orbital localization of ponderomotive forces. 

\begin{figure}[b]
  \centering
  \vspace{-15pt}
  \includegraphics[width=0.7\linewidth]{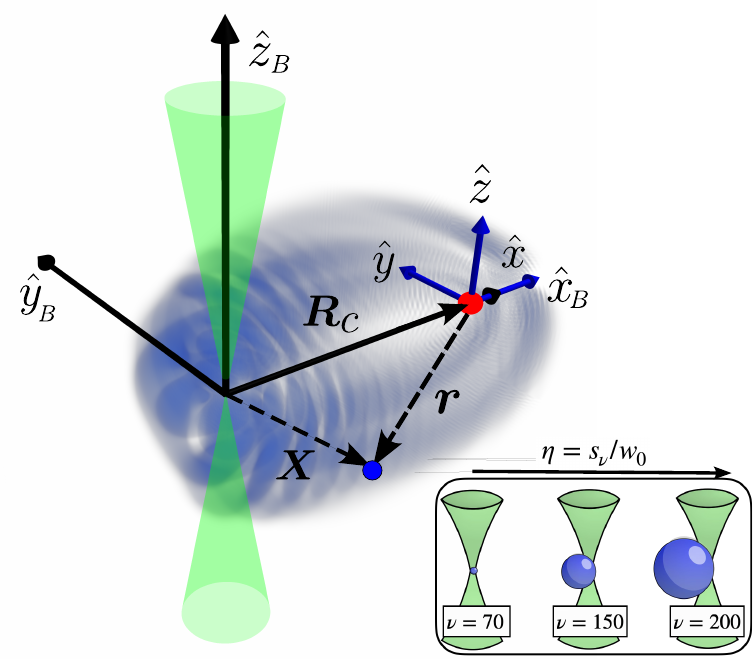}
  \vspace{-5pt}
  \caption{Experimental scheme and coordinate system: A tweezer beam is focused at the origin. The ionic core of the Rydberg atom (red sphere) is located at $\boldsymbol{R}_c$, the Rydberg electron (blue sphere) is located at  $\boldsymbol{X}$. The tightly focused laser beam strongly perturbs the quasi-degenerate Rydberg levels, resulting in localized electronic states reminiscent of the ``trilobite" orbitals in long-range Rydberg molecules (blue density plot). The inset depicts the relative size of a Rydberg atom compared to a tweezer with waist $w_0=480$ nm as the ratio of Rydberg orbit to tweezer waist $\eta$ increases.}
  \label{fig:frames}
\end{figure}

\begin{figure*}[t]
  \centering
\includegraphics[width=\textwidth]{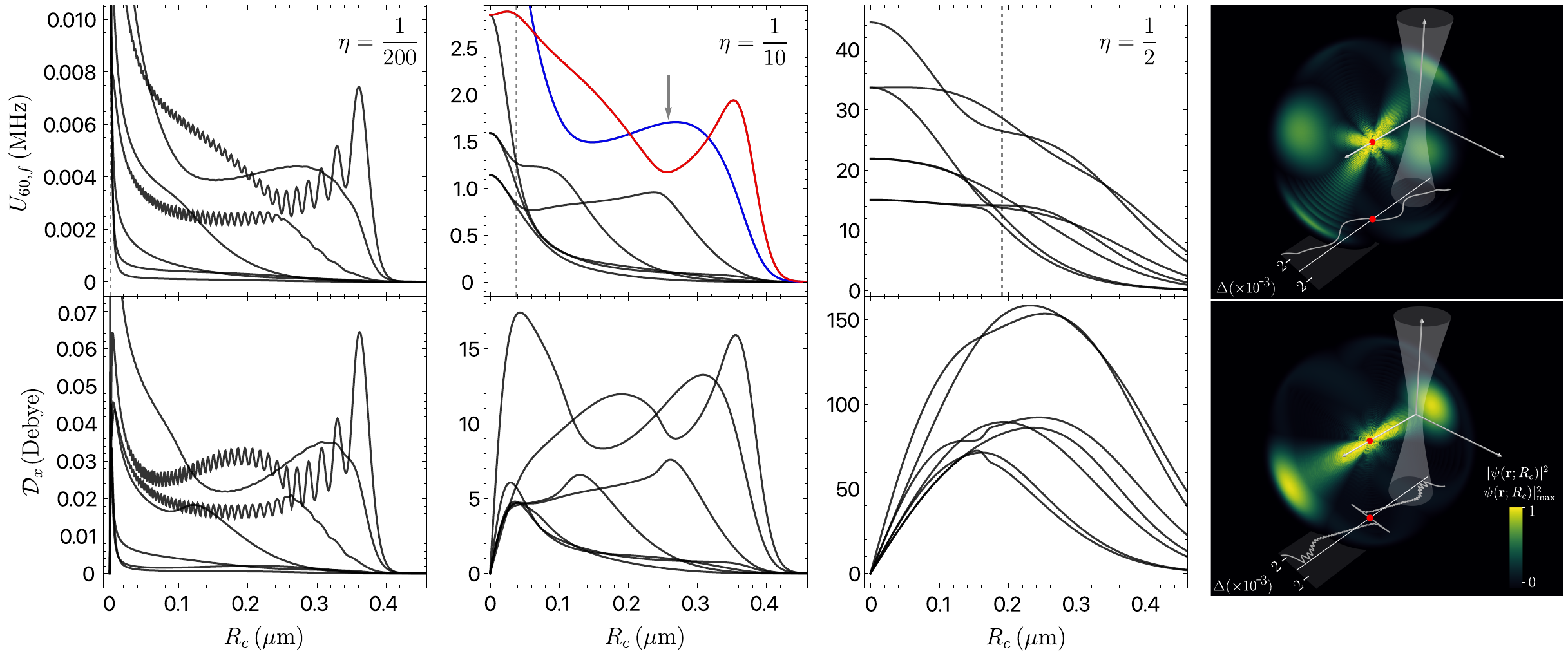}
  \caption{Ponderomotive energies $U_{60, f}$ and electric dipole moment $\mathcal{D}_x=e \langle 
  x \rangle$ of a Rydberg atom with principal quantum number $\nu{=}60$ as a function of the core position $\boldsymbol{R}_c=R_c 
\, \hat{\boldsymbol{x}}_B$, for beam waists $w_0{=} \eta s_{\nu}{=}\eta(2 \nu^2 a_0)$ . As 
  reference, the Rydberg orbital radius is $s_{60}=0.38 \ \mu$m and the gray dashed line indicates $R_c=w_0$. The right column shows the normalized electron density $|\psi(\boldsymbol{r};R_c)|^2/|\psi(\boldsymbol{r};R_c)|^2_{\max}$ on a common $[0,1]$ scale for the eigenstates in the blue (top) and red (bottom) $\eta=1/10$ PECs at core position indicated by the arrow. The white surface represents the 30~\% intensity iso-surface of the beam. To 
  expose the small breaking of inversion symmetry along the $x$ axis, we show in the insets the 
  relative asymmetry in the reduced $x$ density $\rho_x(x)= \int |\psi(\boldsymbol{r})|^2 \, \mathrm{d}y \, 
  \mathrm{d}z$ defined as $\Delta(x)=(\rho_x(x)-\rho_x(-x))/(\rho_x(x)+\rho_x(-x))$ at a given $R_c$.}
  \label{fig:f_block}
\end{figure*}
 
Figure~\ref{fig:frames} illustrates our proposed scheme to sculpt and control the electronic orbital of a Rydberg atom using an optical tweezer passing through the origin of the lab-frame coordinate system, with the ionic core and Rydberg electron lying at $\boldsymbol{R}_c$ and $\boldsymbol{X}=\boldsymbol{R}_c+\boldsymbol{r}$. 
For definiteness, we consider a divalent $^{88}$Sr Rydberg atom and Gaussian beams characterized by waist $w_0$, wavelength $\lambda$, and an effective numerical aperture $\mathrm{NA_{eff}}\coloneqq\frac{\lambda}{\pi\, w_0}\sim 0.3$. 
We consider the tight-focus regime $w_0{=}\eta \, s_\nu$, where $\eta <1$ and $s_\nu \! = \! 2\, \nu^2a_0$ is the characteristic orbital radius of the Rydberg atom with principal quantum number $\nu$.   
The blue density plot in Fig.~\ref{fig:frames} shows a representative electron orbital for the maximally perturbed Rydberg state.
Its strong localization about the tweezer focus and resulting charge separation induces a large dipole moment.

The tweezer's influence on the Rydberg atom is determined by the effective Hamiltonian 
\begin{equation}
\hat H = \hat H_{\mathrm{Ryd}}(\boldsymbol{r}) + \hat V_{\mathrm P}(\boldsymbol{r};\boldsymbol{R}_c,\omega) + \hat V_{\mathrm{core}}(\boldsymbol{R}_c,\omega). 
\label{eq:Ham}
\end{equation}
We assume a single–active-electron picture in which one electron is promoted to a Rydberg state $\ket{\nu \ell m}$, while the second remains part of the core. 
Hence, $\hat H_{\mathrm{Ryd}}$ is the field-free atomic Hamiltonian with eigenenergies $E_{\nu\ell}=-\mathrm{Ry}/(\nu-\mu_{\ell})^2$, with $\mu_\ell$ the quantum defects of $^{88}$Sr, and $\hat V_{\mathrm{core}}$  the frequency-dependent interaction between the ionic core and the light field. 
In the adiabatic approximation, justified by the disparity between typical laser frequencies $\omega$ and both the Rydberg electron’s Kepler frequency and the center-of-mass motion \cite{Duta2000}, $\hat V_{\mathrm P}$ is the static ponderomotive potential. 
These potentials are given by 
\begin{equation}
\hat V_{\mathrm P}(\boldsymbol{r};\boldsymbol{R}_c)= \frac{e^2 \ I(\boldsymbol{X})}{2 m_e\,c\,\varepsilon_0\,\omega^2},
\  \hat V_{\mathrm{core}}(\boldsymbol{R}_c)
= -\frac{\alpha_{\mathrm{ion}}(\omega)}{2\,c\,\varepsilon_0}\,  I(\boldsymbol{R}_c),
\label{eq:Up}
\end{equation}
where $\alpha_{\mathrm{ion}}(\omega)$ is the scalar dynamic dipole polarizability of Sr$^+$ and $I(\boldsymbol{x})$ is the inhomogenous intensity of the beam (see End Matter). 
We consider atomic displacement only in the transverse direction and obtain the adiabatic potential–energy curves (PECs) upon diagonalizing $\hat H$ at each $\boldsymbol{R}_c{=}R_c \, \hat{\boldsymbol{x}}_B$. 
 The core-induced shift $V_{\mathrm{core}}$ is diagonal in the electronic degrees of freedom and can be added \textit{a posteriori}; therefore, we ignore it in most of the following discussion in order to better highlight the role of the Rydberg electron alone. 
 
As the interaction with the tweezer beam is weak at the typical laser power $P_0\!=\!100$ mW, the perturbed electronic states separate into two classes.
The first contains states which, to leading order, still have $\ell$ as a good quantum number due to the inability of the ponderomotive shift to overcome the gap between quantum defect-split energy levels; the second contains strongly perturbed states emerging from the quasi-degenerate high$-\ell$ manifold, taken for Sr to start at $\ell{=}6$, where the quantum defect is effectively zero. Figure~\ref{fig:f_block} presents the PECs $U_{\nu, \ell}(R_c)$ defined with respect to the unperturbed energy $E_{\nu \ell}$ for a representative Rydberg state of the first class having $\nu \!=\!60$ and $\ell\!=\!3$.
 We consider three exemplary $\eta$ values ranging from $\eta=1/200$, which is experimentally unrealistic but theoretically revealing, to $\eta = 1/2$,  well within experimental limits.  
In each case, all $2\ell+1$ PECs are blue-detuned due to the repulsive interaction with the tweezer. As $\eta$ increases, the oscillatory structure in these curves is progressively suppressed while the overall topology remains unchanged. Beyond $\eta\gtrsim 0.25$, the curves become essentially monotonic. 
Additionally, the strength of the perturbation increases with $\eta$.

The tweezer interaction does, for the confined planar case considered here, preserve several reflection symmetries. 
This leads to the existence of real crossings between some pairs of PECs visible in Fig.~\ref{fig:f_block}. 
Moreover, for a fixed $\ell$,  the $2\ell+1$ zero-field degenerate magnetic sub-levels decompose into $\ell$ and $\ell+1$ states, respectively, having odd or even reflection parity with respect to the $xz$ plane.  
The odd parity states are only weakly perturbed in the tight-waist limit. 

The ponderomotive potential admixes states of opposite parity, yielding a nonzero permanent electric dipole in the core frame. 
 The electronic cloud is polarized along $x$; 
accordingly, the bottom row of Fig.~\ref{fig:f_block} reports the dipole moment's $x$-component $ \mathcal{D}_x$.
Because the $\ell$ admixture is small, the symmetry-breaking is not discernible in the real-space densities of Fig.~\ref{fig:f_block}: as shown in the inset, the relative asymmetry $\Delta(x)$ is on the order of $\sim 10^{-3}$, and is positive for positive $x$ consistent with the repulsive effect of the tweezer on the electron. 
Given the large length scales involved, even this small imbalance is sufficient to produce a significant dipole moment for the $60f$ states shown. 

\begin{figure}[t]
  \centering
  \includegraphics[width=\linewidth]{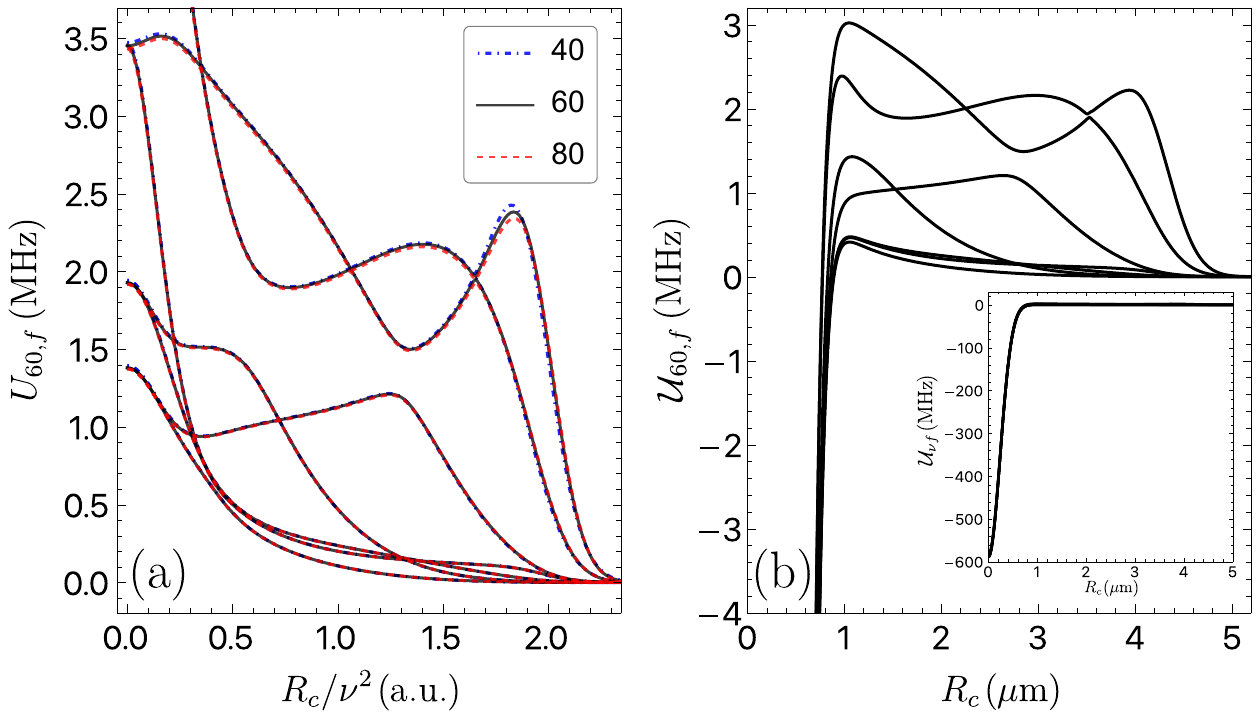}
  \caption{(a) Ponderomotive energies $U_{60, f}$ as function of scaled displacement $R_c/\nu^2$ for $\eta=0.1$ and different $\nu$. (b) Extrapolated total PECs  $\mathcal{U}_{60, f}$ including $\hat V_{\mathrm{core}}$ in the Hamiltonian for $\nu{=}200$  and a tweezer with $\lambda{=}460$ nm and $w_0{=}480$ nm (resulting in $\eta{=}0.11$).  The inset in (b) shows the full scale shifts. The extrapolation is further corrected to account for the weak $\nu-$dependence of the outermost maximum.}
  \label{fig:f_scaling}
\end{figure}

From a semiclassical analysis of the matrix elements of $\hat V_{\mathrm P}$ in the limit $w_0 \! \to \! 0$, we obtain 
\begin{equation}
U^{i}_{\nu, \ell} \sim C_1 P_0 \, \mathrm{NA_{eff}}^2  \eta^2 F^i_\ell \left( R_c/(a_0 \, \nu^2) \right),
\label{eq:scaling_low}
\end{equation}
where $F^i_\ell$ is a dimensionless function that depends on $R_c$ and $\nu$ only through the scaled core displacement $R_c/(a_0  \nu^{2})$ and $C_1=\frac{e^2}{2 m_e\,c^3\,\varepsilon_0}$ \footnote{See End Matter for details and to see how $F_\ell^i$ can be expressed in terms of Elliptic Integrals.}. 
Hence, the energy shifts for the low-$\ell$ states are essentially independent of $\nu$ for fixed $\eta$. 
This universality applies to the envelopes only: when oscillations are present, their fine structure (well depths and minimum locations) retains a residual $\nu$ 
dependence that weakens as $\nu$ increases. 
Figure~\ref{fig:f_scaling}(a) illustrates the scaling, which differs markedly from Rydberg--ground-state atom interactions \cite{DeSalvo2015, Pfau2014}, where the interaction strength decreases as $\nu^{-6}$.
The dipole moment does not behave as universally, as it is primarily a second-order effect with a non-trivial $\nu$ dependence scaling approximately as
$\mathcal{D}_x \! \sim \! P_0 \, \eta^{2}\,\nu^{5}$ at fixed $\mathrm{NA_{eff}}$. 

A key benefit of these scaling laws is that they enable \emph{extrapolation} across principal quantum numbers. Holding $P_0$, $\mathrm{NA}_{\mathrm{eff}}$, and $\eta$ fixed, the same ponderomotive response is obtained for different combinations of  $\nu$, $w_0$, and $\lambda$. 
This is particularly useful for predicting the behavior for large $\nu$ and realistic laser parameters, where the necessary basis size makes calculations intractable, from results obtained for a reference configuration $(\nu',w_0',\lambda')$, as illustrated in Fig.~\ref{fig:f_scaling}(b).
Here, the total $\nu {=} 200$ PECs, including the core contribution, that follows the Gaussian beam profile, are shown. At this large, but experimentally accessible, $\nu$ value, reaching $\eta \! \sim \! 1/10$ with $\mathrm{NA_{eff}\sim0.3}$, requires visible wavelength tweezer parameters $\lambda=460$ nm and $w_0=480$ nm. This beam waist is close to that  reported in Ref.~\cite{Holzl2024} for a tweezer with $\lambda=540$ nm and $w_0=564$ nm. For those parameters one obtains $\eta=0.13$, for which the PECs remain qualitatively unchanged, differing mainly by the $\eta^2$ rescaling. For our illustrative parameter set, using the $\nu$-scaling of $\mathcal{D}_x$, we infer dipole moments as large as a few kilo-Debye for Rydberg $f$-states. Somewhat smaller values are found for the other low-$\ell$  states due to their larger energy gaps from Rydberg states of opposite parity; see End Matter.

The broad local minimum in the second-highest potential curve, with an equilibrium position at about 3 $\mu$m, can support a series of bound states (in this confined 1D system) with a level spacing of around 20 kHz. 
The Rydberg atom would thus be trapped by the optical tweezer, even with a nearly 3 $\mu$m separation between the Rydberg core and the tweezer focus. Unlike previously considered ponderomotive bottle traps \cite{Duta2000,Barredo2020}, here it is the local reshaping of the electron orbit which induces a transverse trapping force which is highly asymmetric with respect to the atom. 

The large MHz-scale energy shifts and splittings we observe readily allow for probing the eigenstates spectroscopically by laser excitation of the Rydberg electron in the presence of the tweezer. 
The dipole moments can be measured via Stark spectroscopy.
In a second scenario, we consider adiabatic modulation of the tweezer intensity after a specific Rydberg state is prepared. Applying moderate magnetic fields allows a single PEC to be energetically isolated, which in turn enables dynamical control of the orbital's dipole moment at MHz-scale bandwidth without detrimental non-adiabatic transfer between the eigenstates. Importantly, we find that the dipole moment is largely preserved in the presence of an additional Zeeman splitting (see End Matter). This results in a giant atomic dipole antenna which can be locally controlled at the single-particle level. Adjacent receiver Rydberg atoms could probe these Hertzian dipoles, for example through radio-frequency dressed resonances \cite{Adams2020}.

\begin{figure}[t]
  \centering
  \includegraphics[width=\linewidth]{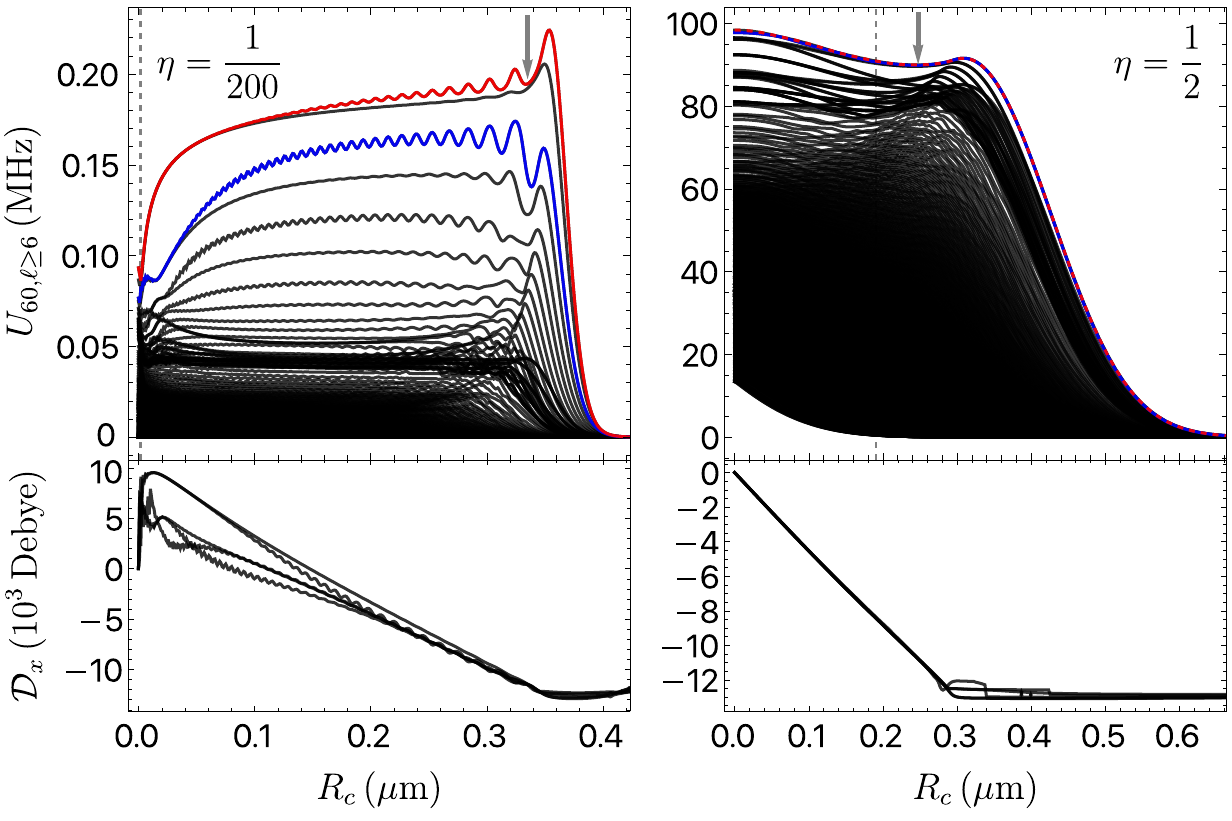}
  \caption{Energies $U_{60, \ell \geq 6}$ and dipole moments for the $\nu=60$ high-$\ell$ manifold as a function of the core position $R_c$ and different reduced waists $\eta$. For the lower part of the manifold, only every fifth state energy is shown. The displayed dipole moments correspond to the five highest energy states.  }
  \label{fig:high_l}
\end{figure}

We now consider the influence of the tweezer on the quasi-degenerate high-$\ell$ hydrogenic manifold. The PECs and selected dipole moments for the two extreme cases $\eta = 1/200$ and $\eta = 1/2$ are shown in Fig.~\ref{fig:high_l}. 
For $\eta=1/2$, all of the unperturbed states are significantly affected by the tweezer, creating a dense continuum. 
In contrast, the tightly focused tweezer in the $\eta = 1/200$ case modifies a handful of levels much more strongly than the rest. 
This has its origin in the approximate separability of the tweezer ponderomotive potential, which in this limit  $w_0 \! \ll  \! s_\nu$ becomes 
sharply localized in the transverse to beam direction and can be approximated by 
\begin{equation}
V_{\mathrm P} \sim (C_1/4) \, P_0 \, \mathrm{NA_{eff}}^2 \, w_0^2  \, \delta(x+R_c) \, \delta(y).   
 \label{eq:Vp_delta}
\end{equation}

Consequently, the three-dimensional intensity-weighted matrix element reduces to a one-dimensional 
overlap integral along the beam axis.
The transverse-$\delta$ reduction implies that the ponderomotive coupling
within the degenerate hydrogenic manifold is effectively low rank. Only a small set of perturbed
linear combinations with substantial axis overlap acquire appreciable shifts, while the majority of orthogonal combinations have negligible overlap and remain only weakly perturbed.
This qualitative structure persists for modest $\eta$: increasing $\eta$ mainly introduces transverse averaging over a radius $\sim ~\! w_0$, which smooths fine modulations, yet preserves the separation between many weakly perturbed states and a  small set of strongly shifted, localized ones. Only for sufficiently large $\eta\sim 1/2$ does the 
perturbation cease to be effectively low rank, and the distinction between perturbed and unperturbed combinations progressively disappears. 
The magnitude of the energy shifts shows a weak $\nu$-dependence at fixed $\eta$: in the tight-waist limit ($\eta \ll 1$) it increases approximately linearly with $\nu$, whereas as $\eta$ approaches unity this trend progressively weakens and the shifts become nearly $\nu$-independent.

\begin{figure}[b]
  \centering
  \includegraphics[width=\linewidth]{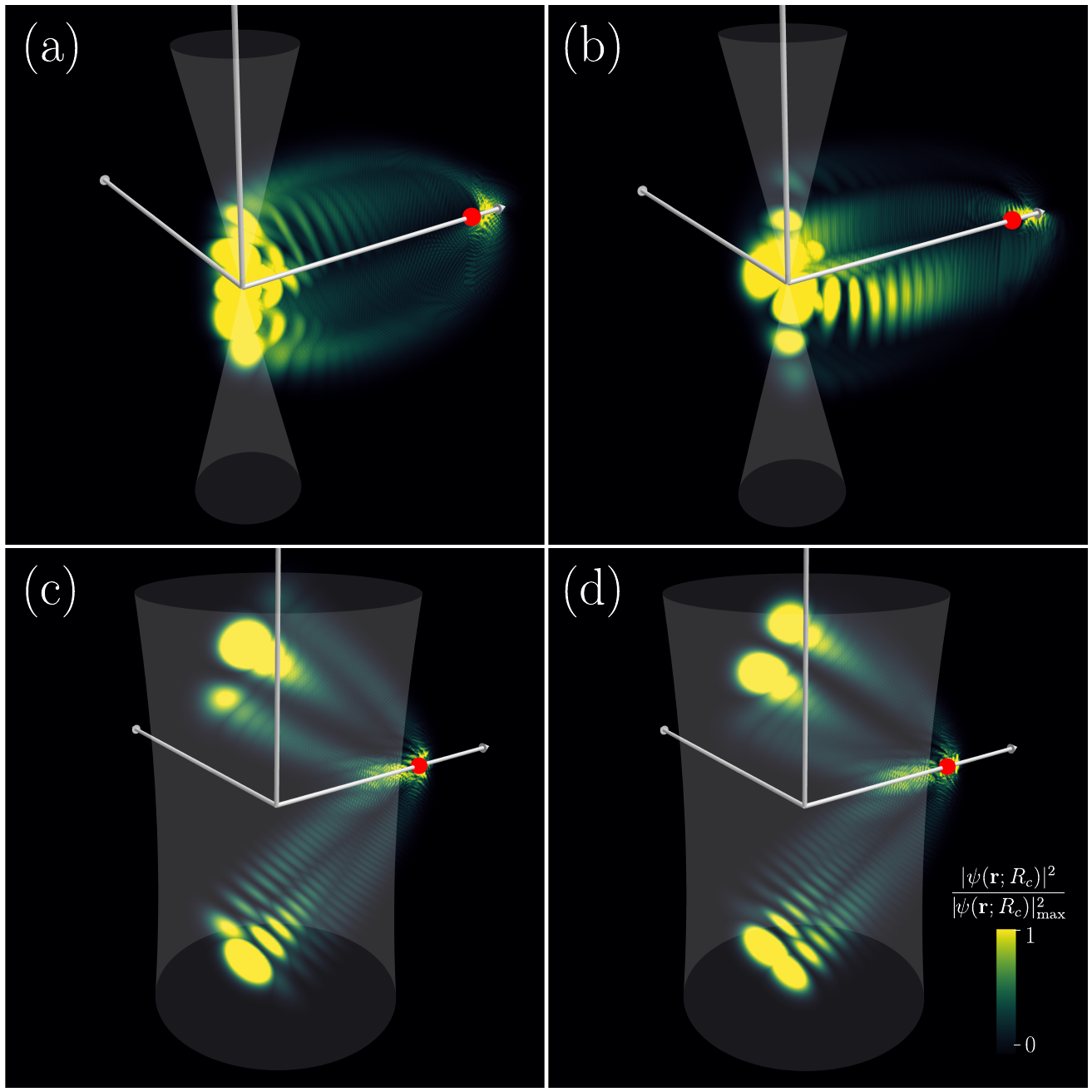}
  \caption{Normalized electron density for the red- and blue-highlighted PECs in Fig.~\ref{fig:high_l} at core position marked by the gray arrow. Panels (a) and (b) correspond to $\eta=1/200$, while panels (c) and (d) correspond to $\eta=1/2$. The beam iso-surface is set to 30 \% of the on-axis intensity.}
  \label{fig:trilobite}
\end{figure}

The shifted eigenstates selected by this axis-localized coupling are 
superpositions of the hydrogenic basis that concentrate electronic weight near the beam axis.
In the strict $w_0 \! \to \! 0$ limit, the highest-energy state approaches 
a pure ``trilobite" state, identical to that produced by a pure delta function potential at $-\boldsymbol{R}_c$ \cite{Greene2000}.  
Such a state thus realizes the ``ghost" molecular bond predicted to arise from carefully tailored time-dependent perturbation of a Rydberg atom \cite{eiles2018theoretical}. 
The highest-energy states for larger $\eta$ in the block remain strongly axis-localized: most of the probability density is confined to a narrow tube around the tweezer axis, with only faint trilobite-like lobes and fringes extending along the classical Kepler ellipse, as shown in Fig.~\ref{fig:trilobite}.
Owing to their strong spatial localization, these eigenstates exhibit very large dipole moments: for $\nu=60$ we obtain values on the order of $1000$~Debye (Fig.~\ref{fig:high_l}), and the dipole magnitude increases further with $\nu$.
For $\eta=1/2$ the electron density is so tightly confined to the axial tube that the 
linear dipole moment in Fig.~\ref{fig:trilobite} is excellently approximated by that of a point charge $0.9 \, e$ at the beam center with the ionic core at $R_c$.
The strongly polar states shown in Fig.~\ref{fig:trilobite} could be optically accessed via microwave coupling from low-$\ell$ states; for $\nu=200$ and $\eta=1/2$, we estimate a Rydberg $g$-state fraction on the order of $0.5$\%. 

To conclude, we have demonstrated that a tightly-focused tweezer can be used for spatio-temporal control of a Rydberg atom via its localized interaction with the electronic orbital.
The Rydberg electron weakly (strongly) localizes away from (near) the tweezer in the case of low-$\ell$ (high-$\ell$) states, leading to the formation of a permanent dipole moment and opportunities to trap the Rydberg atom in a local minimum of the position-dependent energy shift. Although high-NA tweezers ($\mathrm{NA} \approx 0.5$) are now standard for single atom trapping \cite{Endres2016,Barredo2016} including alkaline-earth Rydberg atoms \cite{Holzl2024,Wilson2022}, we adopt a more modest numerical aperture to remain close to typical experimental conditions. Consequently, for realistic laser frequencies the tight-waist regime is reached only at very large $\nu$ (typically $\nu \gtrsim 150$), where the Rydberg orbit is sufficiently extended. 
Rigorous scaling laws allow for easy extraction of their properties from calculations at lower $\nu$. Positional averaging is irrelevant in this regime. Fluctuations of the relative core-tweezer arise both from the core motion on the ground state of the tweezer, with a typical oscillator length $\sim 30$ nm, and from beam pointing fluctuations. For large-$\nu$, these fluctuations are small compared with the $>1\mu$m scale over which the PECs and dipole moments vary. In addition, the corresponding Rydberg-state lifetimes ($\sim 100$ $\mu$s) are expected to exceed by a wide margin the timescales associated with the MHz dynamics considered here. Following the local-intensity argument of Refs.~\cite{Younge_2010,Cardman_2021}, we have also verified  that photoionization is negligible in the regimes considered here and it is further suppressed for the relevant configurations where the core lies outside the high-intensity region of the tweezer. To prepare the initial Rydberg atom in the appropriate configuration for the tweezer, it could be trapped initially using a low-intensity tweezer acting on its core electron, a configuration potentially enabling coupled dynamics between the tweezer-trap formed by the Rydberg electron and the tweezer trapping the ionic core \cite{pendse2025transferring}.

Future directions include freeing the ionic core to move along the tweezer axis direction, which would break the reflection symmetry and open couplings that are symmetry-forbidden in the present configuration. Along the same line, using spatially structured light fields, e.g. beams carrying orbital angular momentum \cite{Rodrigues2016, Mukherjee2018} or controlled beam deformations, could introduce additional anisotropies to access higher-$\ell$ states of ultralong-range Rydberg molecules \cite{Hamilton_2002, Giannakeas2020}.
 The highly dipolar states could be excited in spatially structured tweezer arrays to utilize the anisotropy of the dipole-dipole interaction \cite{Eiles2017,Rivera2021}. 
 Finally, the presence of multiple tweezers would provide an alternative path towards the concept of a Rydberg composite -- the many-perturber limit of a polyatomic Rydberg molecule \cite{Hunter2020,Eiles2023,Eiles2024}. 

\textit{Acknowledgments:} F.M. acknowledges funding
from the Federal Ministry of Research, Technology and
Space under the Grants CiRQus and QRydDemo, and
the Horizon Europe Programme HORIZON-CL4-2021-
DIGITAL-EMERGING-01-30 via Project No. 101070144
(EuRyQa).  T.P. acknowledges funding from the European Research Council (ERC)(Grant Agreement No. 101019739)

% ===========================
\clearpage
\appendix
\makeatletter
\section{End Matter}
\section{Details of the ponderomotive potential}
\label{app:appendA}
The intensity profile for a Gaussian beam is 
\begin{align}
I(\boldsymbol{r}) \;=\; \frac{2P_0}{\pi\,w(z)^2}\,
\exp\ \left[-\,\frac{2r_{\perp}^2}{w(z)^2}\right],
\label{eq:gaussian} \\
w(z)=w_0\sqrt{1+\left(\frac{z}{z_R}\right)^2}, \ \ 
z_R=\frac{\pi\,w_0^2}{\lambda}, \nonumber
\end{align}
with transverse radius $r_{\perp}$ and Rayleigh range $z_R$. In the hydrogenic spherical basis $\{\psi_{\nu\ell m}(\boldsymbol{r})\}$ the ponderomotive matrix elements are, with $i=\{\nu \ell m\}$,
\begin{align}
(V_P)_{ii'} & = 
\frac{C_1 c^2}{\omega^2} 
\int & \mathrm{d}^3\boldsymbol{r}\;\psi_{\nu\ell m}^*(\boldsymbol{r})\,I(\boldsymbol{R}+\boldsymbol{r})\,\psi_{\nu'\ell' m'}(\boldsymbol{r}).
\label{eq:Up-matrix}
\end{align}
The azimuthal $\varphi$ integral can be performed analytically, leaving 
\begin{equation}
    G_{w_0}(r_{\perp}; z, R_c)=\frac{1}{\xi^2} \exp \left[ -2 \, \frac{r_{\perp}^2+R_c^2}{\xi^2} \right] I_M \left( -\frac{4 \, R_c \, r_{\perp}}{\xi^2}\right)
    \label{eq:bessel}
\end{equation}
as the kernel for the remaining two-dimensional integral, where
 $M=m'-m$, $I_M(x)$ the modified Bessel function of the first kind and $\xi^2=w_0^2(1+(z/z_R)^2)$.
\section{Numerical details}

We use numerically obtained wave functions for the singlet states of $^{88}$Sr Rydberg states $\psi_{\nu l m}$ including quantum defects up to $\ell_{\mathrm{maxQD}}=5$ \cite{Patsch2021}. The laser power is fixed at $P_0=~100~\mathrm{mW}$. The selected $\mathrm{NA_{eff}}= 0.3$ is well within the paraxial regime. Numerical diagonalization is restricted to a basis including the full target $\nu-$manifold and the quantum defect states of the adjacent manifolds. It suffices to consider only states within the target $\nu-$manifold to reach energy convergence, whereas converging other observables such as dipole moments requires including the neighboring manifolds $\nu \pm1$. Including all $l$ and $m$ in the manifolds gives a basis size $\nu^2+2(\ell_{\mathrm{maxQD}}+1)^2$ for a target $\nu-$manifold. The core potential is calculated using $^{88}$Sr data from Ref. \cite{Kiruga2025}.

\section{Tight-waist limit}
In the limit $w_0  \! \to \! 0$, the argument of the Bessel function in Eq.~\ref{eq:bessel} becomes large for any $R_c$ not close to zero. In this asymptotic regime, the kernel is sharply localized in the transverse coordinate $r_\perp$, effectively approaching a one-dimensional $\delta$-function (up to an overall normalization factor and once the cylindrical volume element is included). Moreover, since in the core frame the beam center lies at $\varphi=\pi$, the ponderomotive potential matrix element simplifies to 
\begin{align}
\label{eq:axis_overlap}
(V_P)_{ii'} &=B(w_0)\Big[ 
\int f_{ii'}(z) \, \mathrm{d}z 
\\&+ \int  \frac{w^2(z)}{8} \! \left(\!\partial_{r_{\perp}}^2\!+\!\frac{\partial_{r_{\perp}}} 
{R_c} \!-\!\frac{M^2}{R_c^2}\right)  f_{ii'}(z) \, \mathrm{d}z \Big],
\nonumber
\end{align}
where  $f_{ii'}(z)=\psi_{\nu'\ell' m'}(-\!R_c,\!0,\!z)\psi_{\nu'\ell' m'}(-\!R_c,\!0,\!z)$ and $B(w_0)=\frac{P_0 C_1 c^2}{\omega(w_0)^2}=(C_1/4)P_0 \mathrm{NA_{eff}}^2 w_0^2$. The leading term of Eq.~\eqref{eq:axis_overlap} gives the effective potential of Eq.~\eqref{eq:Vp_delta}. 

Using the symmetry of $\psi_{\nu \ell m}(\varphi\!=\!\pi)$ under $m \!\to \! -m$, the interaction matrix can be diagonalized exactly within each low-$\ell$ block by working with real combinations of $m$ states defined, for  $m{>}0$, as $\ket{c_m}\coloneqq(\ket{m}+(-1)^m\ket{-m})/\sqrt2 \propto \cos(m\varphi)$ and $\ket{s_m}\coloneqq  \, (\ket{m}-(-1)^m\ket{-m})/\sqrt2 \propto \sin(m\varphi)$ (with $\ket{0}$ unchanged). The sets $\lbrace \ket {c_m}\rbrace$ and $\lbrace \ket {s_m}\rbrace$ decouple identically as a consequence of the reflection symmetry. Moreover, since $\sin(m\varphi){=}0$ on-axis,  the $\lbrace \ket{s_m} \rbrace$ set has vanishing axis overlap and yields $\ell$ null eigenvalues 
within each $\ell$ block.

\begin{figure}[t]
  \centering
  \includegraphics[width=\linewidth]{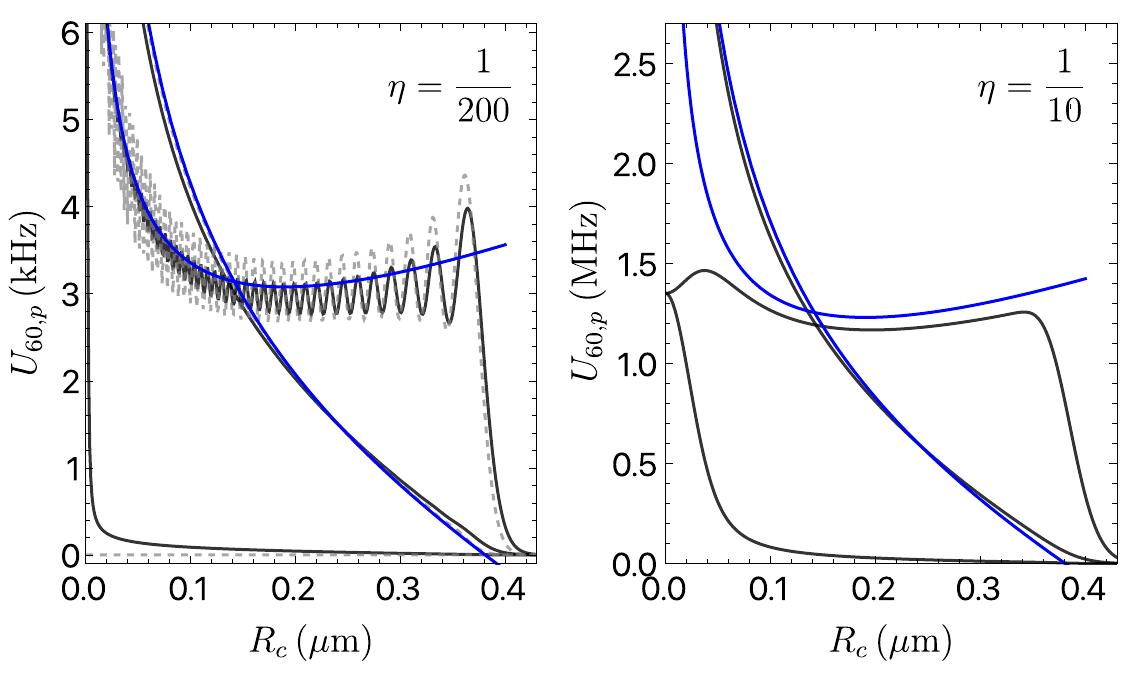}
  \caption{PECs for the $p-$block obtained using three-dimensional integration (black continuous), one-dimensional overlap Eq.~\eqref{eq:1dintegral} (gray dashed) and semiclassical envelope Eq.~\eqref{eq:envelope} (blue continuous) for two waist parameters $\eta$.}
  \label{fig:comparison}
\end{figure}

We illustrate diagonalization explicitly for the $\ell=1$ block. The $\lbrace \ket{s_1},\ket 0\, \ket{c_1} \rbrace$ basis is  already decoupled from only $m$-parity considerations. The  eigenvectors are $\ket{s_1}~=~(\ket 1+~\ket{-1})/\sqrt{2}$ with energy $\tilde{U}^{(1)}_{\nu,p}= \!0$, $\ket 0$ and $\ket{c_1}=(\ket 1-\ket{-1})/\sqrt{2}$ with respective energies
\begin{align}
\tilde{U}^{(2)}_{\nu,p}(R_c)&= B(w_0) \int \left| \psi_{\nu10}(-R_c,0,z) \right|^2 \mathrm{d}z, \nonumber \\ 
\tilde{U}^{(3)}_{\nu,p}(R_c)&= 2 \, B(w_0) \int \left| \psi_{\nu11}(-R_c,0,z) \right|^2 \mathrm{d}z.
\label{eq:1dintegral}
\end{align}

Semiclassical evaluation yields closed-form approximations for the envelope of these PECs in terms of complete 
elliptical integrals of the first and second kind $K$ and $E$, with elliptical parameter $\chi=1/2-R_c/(4 a_0\nu^2)$, valid within the classically allowed region $R_c<2 a_0 \nu^2$. From these expressions the universal scaling given by Eq.~\eqref{eq:scaling_low} is implied. In this particular $\ell=1$ case,

\begin{align}
\tilde{U}^{(2)}_{\nu,p}(R_c)& \sim  \frac{2 B(w_0)/a_0^2}{\pi^2 \nu^3 \sqrt{R_c/a_0}} \left[(1-\chi) K(\chi) -(1-2 \chi)E(\chi) \right], \nonumber\\ 
\tilde{U}^{(3)}_{\nu,p}(R_c)&\sim \frac{B(w_0)/a_0^2}{2 \pi^2 \nu^3 \sqrt{R_c/a_0}} \left[(4\chi-1) K(\chi) +4(1-2\chi)E(\chi) \right].
\label{eq:envelope}
\end{align}
In Fig.~\ref{fig:comparison} we compare the numerical three-dimensional calculation with the one-dimensional overlap model.

\section{Adiabatic state preparation and dipole moment modulation}

\begin{figure}[b]
  \centering
  \includegraphics[width=0.9\linewidth]{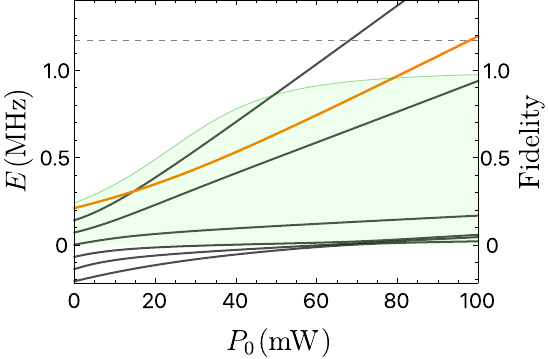}
  \caption{Energy splitting evolution while ramping $P_0$ in a weak magnetic field $B=50$ mG for the situation $\nu=60$ and $\eta=1/10$. The energy of the state that adiabatically connects to the target state is shown with the orange line. The fidelity between the produced state at each $P_0$ and target state is showed as the green shaded region.}
  \label{fig:Pramp}
\end{figure}

\begin{figure}[t]
  \centering
  \includegraphics[width=\linewidth]{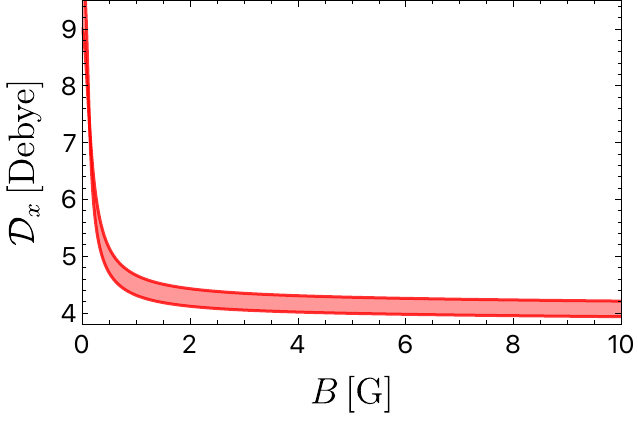}
  \caption{Dipole moment of the perturbed $\nu=60$ $f-$state for $\eta=1/10$ as a function of the applied magnetic field $B$. The shaded region shows the variation of $\mathcal{D}_x(B)$ for $R_c$ around the equilibrium position.}
  \label{fig:dipolevsB}
\end{figure}

One route to prepare the polarized low angular momentum states is to start from a well-defined $m$ level of an $f$-state Rydberg atom excited in a magnetic field $B$ without the tweezer present, and then adiabatically ramping on the tweezer intensity. Figure~\ref{fig:Pramp} shows how the level splittings evolve with the tweezer power $P_0$ in a $50$mG field. 
The state corresponding to the minimum of the red $\eta=1/10$ PEC in Fig.~\ref{fig:f_block} adiabatically connects to the zero-field $m=3$ state. The shaded region indicates the fidelity between the polarized state obtained at given $P_0$ and the target state in Fig.~\ref{fig:f_block} computed for $B=0$. Note that there is a real crossing at $\sim 16$ mW due to the reflection symmetries discussed above; the other curve starts from an $m=2$ level. 

The dipole moment of the perturbed low$-\ell$ states is largely preserved as the magnetic field increases.  Figure~\ref{fig:dipolevsB} shows $\mathcal{D}_x$ of the perturbed $f-$state at core positions around the zero-field PEC minimum. As $B$ increases, the dipole moment saturates to a nearly constant value about half its zero-field value. At such larger $B$-fields single PECs can be energetically isolated, which enables the fast modulation of the dipole moment as discussed in the main text. This persists until the Zeeman splitting becomes large enough to bring the $f$ and $g$ blocks into near-degeneracy, where additional resonant mixing can strongly modify the state's character. Figs.~\ref{fig:Pramp} and ~\ref{fig:dipolevsB} further illustrate that the induced dipole moment is robust against small variations on $P_0$ and $R_c$.

\section{Dipole moments of other $\ell$ levels}
\begin{figure}[!h]
  \centering
  \includegraphics[width=\linewidth]{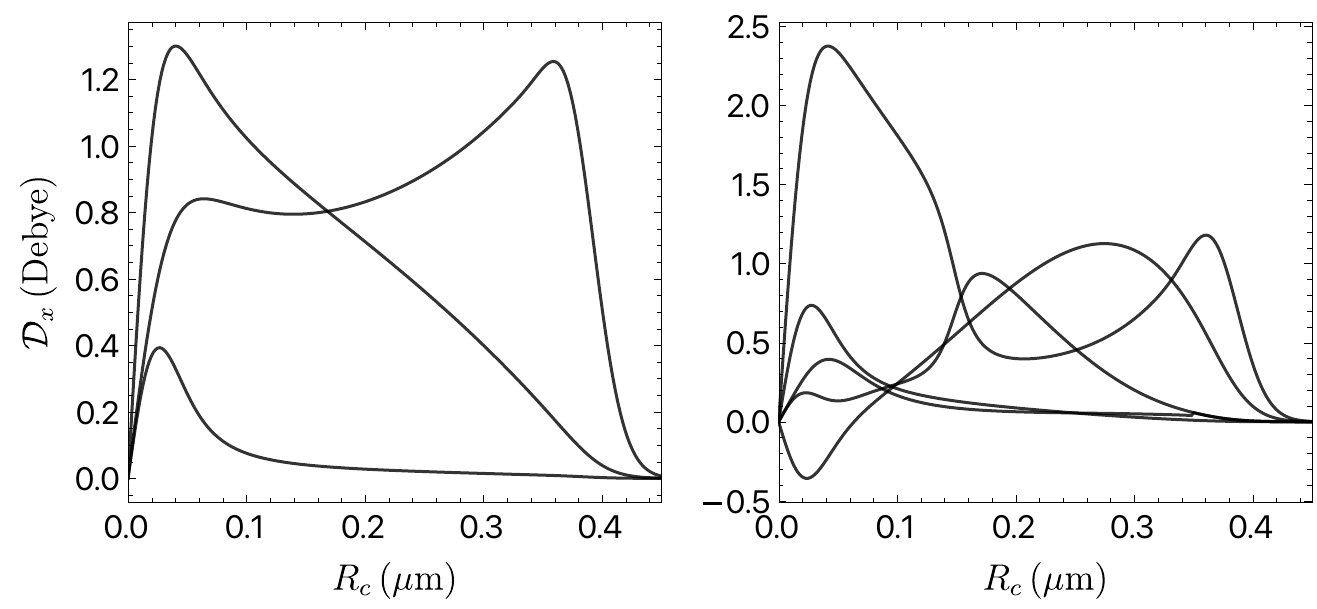}
  \caption{Dipole moments for tweezer-perturbed Rydberg $\nu=60$ $p-$ and $d-$states with $\eta=1/10$.}
  \label{fig:dipole_low}
\end{figure}

The dipole moments of the low$-\ell$  quantum defect-states are smaller than those of the $f-$states, yet remain appreciable for larger $\nu$ as illustrated in Fig.~\ref{fig:dipole_low} for $\nu=60$. Extrapolating with the approximated $\nu$-scaling to $\nu=200$, we estimate dipole moments of the order of $10^2$ Debye even for these levels. These dipole moments are sensitive to the quantum defects, and other atomic species could be used to increase the size of certain low-$\ell$ states. 

\clearpage

\end{document}